\documentclass[journal abbreviation]{copernicus}

\begin{document}

\title{Relation between current sheets and vortex sheets in
stationary incompressible MHD}
\author[1,2]{Dieter Nickeler}
\author[2]{Thomas Wiegelmann}
\affil[1]{Astronomical Institute AV \v{C}R Ond\v{r}ejov, Fri\v{c}ova 298, 25165 Ondrejov, Czech Republic}
\affil[2]{Max-Planck-Institut f\"ur Sonnensystemforschung,
Max-Planck-Strasse 2, 37191 Katlenburg-Lindau, Germany}

%% The [] brackets identify the author to the corresponding affiliation, 1, 2, 3, etc. should be inserted.

\runningtitle{Current sheets and vortex sheets}
\runningauthor{Nickeler \& Wiegelmann}
\correspondence{Nickeler\\ (EMAIL: nickeler@asu.cas.cz)}

\received{}
\pubdiscuss{} %% only important for two-stage journals
\revised{}
\accepted{}
\published{}

%% These dates will be inserted by the Publication Production Office during the typesetting process.

\firstpage{1}

\maketitle

\begin{abstract}
Magnetohydrodynamic configurations with strong localized current concentrations
and vortices play an important role for the dissipation of energy in
space and astrophysical plasma. Within this work we investigate the
relation between current sheets and vortex sheets in incompressible,
stationary equilibria. For this approach it is helpful that the similar
mathematical structure of magnetohydrostatics and stationary incompressible
hydrodynamics allows us to transform static equilibria into stationary ones.
The main control function for such a transformation is the profile of the
Alfv\'en-Mach number $M_A$, which is always constant along magnetic field
lines, but can change from one field line to another. In the case of a
global constant $M_A$, vortices and electric current concentrations
are parallel. More interesting is the nonlinear case, where $M_A$ varies
perpendicular to the field lines. This is a typical situation at boundary
layers like the magnetopause, heliopause, the solar wind flowing around
helmet streamers and at the boundary of solar coronal holes.
The corresponding current and vortex sheets show in some cases also an alignment, 
but not in every case. For special density distributions in 2D it is possible to have
current but no vortex sheets. In 2D vortex sheets of field aligned-flows
can also exist without strong current sheets, taking the limit
of small Alfv\'en Mach numbers into account. The current sheet can vanish if
the Alfv\'en Mach number is (almost) constant and the density gradient is large
across some boundary layer. It should be emphasized that the
used theory is not only valid for small Alfv\'en Mach numbers $M_A\ll 1$, but also for
$M_A\stackrel{<}{\sim} 1$. Connection to other theoretical approaches and observations and
physical effects in space plasmas are presented. Differences in the various aspects of theoretical
investigations of current sheets and vortex sheets are given.
\end{abstract}

%% only used for copernicus2.cls
%\abstract{
%\keywords{TEXT}}

\introduction
\label{sec1}
Many structures in magnetospheres or in the solar corona are often described by
quasi-magnetohydrostatic sequences of equilibria, e.g., \cite{1999ESASP.448..871R},
\cite{1982JGR....87.2263S}, \cite{becker:etal01} or \cite{wiegelmann:etal95}.
Current sheets 
are important for storage and release, i.e. dissipation of energy, leading, e.g.,
to magnetic reconnection, see \cite{2005ESASP.600E..31N}, or \cite{2003Natur.425..692S} for
observational aspects of coronal heating.
The formation of thin current sheets, is often done within the frame of
the quasi-static approach, 
%see,%e.g., \cite{1982JGR....87.2263S} or \cite{wiegelmann:etal95},
where the time dependent pressure on the boundary is prescribed.
Via the law of adiabatic change of the flux tubes in the frame of ideal MHD magnetospheric
convection, the flux transport takes place, and forms highly structured current sheets.

Another approach of
the formation of current sheets is a
parametric method, performed, e.g. in \cite{2006A&A...454..797N}
and \cite{2010AnGeo..28.1523N}. Here the parameter is the Alfv\'en Mach number instead of the
control parameters like, e.g., in \cite{1995ApJ...446..377F} and \cite{1999ESASP.448..871R}.
These parameters describe constraints or boundary conditions, like magnetic shear, pressure
or currents.
 
Instabilities of current and vortex sheets, although with very
high Alfv\'en Mach number, have been analysed in \cite{2003PhPl...10.4661B}. In this
article grid-adaptive simulations show that even in very weak magnetic field regimes
($M_A \simeq 30$), the large-scale Kelvin-Helmholtz coalescence process can trigger tearing-type
reconnection events previously identified in cospatial current-vortex sheets.

The paper of \cite{2006PhyD..223...82E} explains how for discontinuous fields a part of the magnetic
flux can slip through the plasma by an averaged turbulent statistic fluctuation
of a {\it subscale electromotive force (EMF)}, i.e. an electric field in the
comoving systems of the ions. Another important aspect of this paper is that current sheets
and vortex sheets both have to exist and \lq intersect in sets of large enough dimensions\rq~to
get \lq ideal\rq~magnetic reconnection, i.e. \lq magnetic flux conservation may be broken in ideal
MHD by nonlinear effects\rq.
Such tangential discontinuities are basically current-vortex sheets. 

From numerical incompressible time dependent
adaptive MHD-simulations by \cite{1998PhPl....5.2544G,2000PhRvL..84.4850G},
it is shown that vortex-sheets align with current sheets. The initial
conditions in this investigation are given by the so called Orszag-Tang vortex.

In the present paper we analyse and explain some of the physical implications of the non-canonical
transformation technique
\citep[as developed and discussed by, e.g.][]{gebhardt:etal92,petrie:etal99} to calculate
steady-state ideal MHD flows with current sheets.
We analyse the feedback-circuits
on magnetohydrodynamic forces that arise, enabling a relaxed, incompressible
steady-state plasma flow along a magnetohydrostatic background field.
We analyse the connection between vortex and current sheets in the frame of this
mathematical technique, but also for general field-aligned flows.
% and
%the connection to \lq ideal\rq~magnetic reconnection
%\cite{2006PhyD..223...82E}.

In section \ref{sec2} we discuss the basic stationary MHD-equations
and introduce a general solution for the incompressible field aligned case,
based on magnetohydrostatic equilibria.
In particular, we investigate the forces induced by the flow.
Section \ref{sec3} contains special cases of initial potential and force-free
equilibria. In section \ref{sec4} we investigate the relation between
electric currents and flow vortices in general and provide analytic
example solutions in section \ref{sec5}. Finally, we discuss the impact
of this work on physical objects like planetary magnetospheres and
the solar coronal plasma.

%% \introduction[modified heading if necessary]

\section{Transformation from magnetohydrostatic equilibria to stationary
MHD configurations.}
\label{sec2}
Under the assumption of a finite Alfv\'en Mach number
the equations of stationary incompressible and field-aligned
MHD are given by
%%%%%%%%%%%%%%%%%%%%%%%%%%%%%%%%%%%%%%%%%%%%%%%%%%%%%
\begin{eqnarray}
\vec\nabla\cdot\vec v &=& 0\, ,\quad\vec v\cdot\vec\nabla\rho = 0 \, ,\label{konti}\\
\rho\left(\vec v\cdot\vec\nabla\right)\vec v &=&\vec j\times\vec B - \vec\nabla p\, ,
\label{euler0} \\
\vec v &=& \frac{M_{A}\vec B}{\sqrt{\mu_{0}\,\rho}} \, , \label{ohm}\\
%\vec v\times\vec B=\vec 0 \label{paral}\\
%\vec\nabla\cdot\vec v=0\label{konti2}\\
\vec\nabla\cdot\vec B &=& 0\label{divb} \, ,
\end{eqnarray}
%%%%%%%%%%%%%%%%%%%%%%%%%%%%%%%%%%%%%%%%%%%%%%%%%%%%%
where $\vec v$ is the velocity, $\rho$ the mass density, $\vec j=\frac{1}{\mu_0}
\vec\nabla\times\vec B$ the current density vector, $p$ the scalar plasma pressure,
and $M_{A}$ is the Alfv\'en Mach number.
%and $\vec E$ the electric field, with the additional constraint for the Ohm's law,
%Eq.\,(\ref{paral}).
The assumption that the magnetic field and the flow are parallel leads
to the conclusion that the electric field has to vanish everywhere identically to
fulfill ideal Ohm's law.

Within this work we solve the equations by an transformation-approach
developed in \cite{gebhardt:etal92}, which transforms a static
solution into a stationary one.
We show that the general solutions of the stationary,
field-aligned incompressible ideal MHD equations lead to a
very close connection between current and vortex sheets.
We also show how the forces are compensated by the used
exact solution method.

In the following we transform the force terms to enlighten
the connection to the magnetic and plasma force terms.
Under the assumption of field aligned stationary incompressible
plasma flow we get the transformation equations
\citep[see][for details]{2010AnGeo..28.1523N, 2006A&A...454..797N}.
Keeping in mind that the mass continuity
Eqs.\,(\ref{konti}), and (\ref{divb}) for every Alfv\'en Mach number $M_{A}$ and
the density $\rho$ satisfying $\vec B\cdot\vec\nabla M_{A}=0$ and
$\vec B\cdot\vec\nabla\rho=0$ is a necessary condition,
the general solution of the system Eqs.\,(\ref{konti})--(\ref{divb}) in 3D, but also in 2D,
is given by
%%%%%%%%%%%%%%%%%%%%%%%%%%%%%%%%%%%%%%%%%%%%%%%%%%%%%%%%%%%%%%%%%%%%%%%%
\begin{eqnarray}
\vec B &&=\frac{\vec B_{S}}{\sqrt{1-M_{A}^2}}\, ,\label{magtrafo} \\
p &&=p_{S} - \frac{1}{2\mu_{0}}\frac{M_{A}^2\left|\vec B_{S}\right|^2}{1-M_{A}^2}
\, , \label{pressuretrafo}\\
%\vec w &&=
\sqrt{\rho} \vec v &&=\frac{1}{\sqrt{\mu_{0}}}\,
\frac{M_{A}\vec B_{S}}{\sqrt{1-M_{A}^2}}
%\equiv\frac{M_{A}\vec B}{\sqrt{\mu_{0}\,\rho}}
\label{streaming2} \, , \\
\vec j &&=\frac{M_{A}}{\mu_{0}}\frac{\vec\nabla M_{A}\times\vec B_{S}}{\left(1-M_{A}^2
\right)^{\frac{3}{2}}}
+\frac{\vec j_{S}}{\left(1-M_{A}^2\right)^{\frac{1}{2}}} \, , \\
\vec\nabla p_{S} &&=\vec j_{S}\times\vec B_{S} \label{mhs1}
\,\, ,
\end{eqnarray}
%%%%%%%%%%%%%%%%%%%%%%%%%%%%%%%%%%%%%%%%%%%%%%%%%%%%%%%%%%%%%%%%%%%%%%%
where the subscript $S$ defines the original magnetohydrostatic fields.
%with $\vec\ w$ as the streaming vector. The streaming vector has a specific physical
%meaning. It is the vector parallel to the plasma flow, and its absolute value
%is the doubled value of the kinetic energy density, i.e. the stagnation pressure,
%%%%%%%%%%%%%%%%%%%%%%%%%%%%%%%%%%%%%%
%\begin{eqnarray}
%\varrho_{\textrm{kin}}=\frac{\vec w^2}{2}=\frac{M_{A}^2}{1-M_{A}^2}\,
%\frac{\vec B_{S}^2}{2\mu_{0}\rho} \,\,\, .
%\end{eqnarray}
%%%%%%%%%%%%%%%%%%%%%%%%%%%%%%%%%%%%%%%

Thus one has a recipe to construct field-aligned, incompressible
flows along magnetohydrostatic structures: the first step is to find a
magnetohydrostatic (MHS) equilibrium, which we label with the subscript
{\it S} for static ($p_{S},\vec B_{S}$). For the mathematical method
introduced by \cite{gebhardt:etal92} the static equilibria is written
with the help of Euler potentials $(\alpha, \beta)$:
$\vec B = \nabla \alpha \times \nabla \beta$ or, if the equilibrium has some
sort of symmetry (e.g. in z-direction) with the flux function
$A$: $\vec B = \nabla A(x,y) \times \vec e_z + B_z(x,y) \vec e_z$.
The next step is to specify a Mach number profile
$M_A(\alpha, \beta)$, or $M_{A}(A)$ in 2D, depending on the Euler potentials or flux
function, respectively. One should be aware that the function
$\rho$ as a function of the two Euler-potential is independent of the choice
of the Alfv\'en Mach number. They are basically two independent functions.
%We will later on see, that this assumption is invalid in special cases for
%specific constraints, e.g. if the MHD flow has no vortices.

The analysis of the forces deals with all the force terms in the Euler-equation of
ideal MHD, namely
%%%%%%%%%%%%%%%%%%%%%%%%%%%%%%%%%
\begin{eqnarray}
\rho\left(\vec v\cdot\vec\nabla\right)\vec v=\vec j\times\vec B - \vec\nabla p
\,\,\, .
\label{euler1}
\end{eqnarray}
%%%%%%%%%%%%%%%%%%%%%%%%%%%%%%%%%%%
The identities Eqs.\,(\ref{magtrafo})--(\ref{streaming2})
%
%\begin{eqnarray}
%\vec w &=& \sqrt\rho\,\vec v
%\label{streaming1}\\
%\begin{eqnarray}
%\vec w &=& \frac{M_{A}\vec B}{\sqrt{\mu_{0}}}\label{streaming2}
%\end{eqnarray}
%
lead with the Eqs.\,(\ref{konti})--(\ref{divb})
%with the help of the Weber-transformation, Eq.\,(\ref{weber}), 
to
%%%%%%%%%%%%%%%%%%%%%%%%%%%%%%%%%%%%%%%%%%%%%%%%%%%%%%%%%%%%%%%%%%%%%%%%
\begin{eqnarray}
 \rho\left(\vec v\cdot\vec\nabla\right)\vec v = 
%\rho\,\vec\nabla\,\frac{\vec v ^2}{2}
%-\rho \vec v\times\vec\nabla\times\vec v\label{weber}\\
% = &&- \vec w\times\vec\nabla\times\vec w+ \vec\nabla\,\frac{\vec w ^2}{2}\\
% = &&-\frac{M_{A} \vec B^2}{\mu_{0}}\,\vec\nabla M_{A} + \frac{M_{A}^2}{\mu_{0}}\,
%\left(\vec\nabla\times\vec B\right)\times\vec B\nonumber\\
%&& + \frac{M_{A} \vec B^2}{\mu_{0}}\,\vec\nabla M_{A} + \frac{M_{A}^2}{2\mu_{0}}
%\vec\nabla\vec B^2\,
%\nonumber\\
%= &&\frac{M_{A}^2}{\mu_{0}}\,\left(\vec B\cdot\vec\nabla\right)\vec B\nonumber\\
%\equiv &&
\frac{M_{A}^2}{\mu_{0}\left(1-M_{A}^2\right)}\left(\vec B_{s}\cdot\vec\nabla\right)\vec B_{s}\,
\label{weber2} .
\end{eqnarray}
%%%%%%%%%%%%%%%%%%%%%%%%%%%%%%%%%%%%%%%%%%%%%%%%%%%%%%%%%%%%%%%%%%%%%%%%%
%Here we made use of the relations Eqs.\,(\ref{konti}), (\ref{ohm}), and (\ref{divb}).
%being derived in the appendix, see
%Eq.\,(\ref{infor1}).
%
This means that the new inertial force, induced by the flow, depends
mainly on the magnetic tension of the magnetohydrostatic field, and of
course on a function of the Alfv\'en Mach number, i.e.
%For small Alfv\'en Mach numbers $M_A \ll 1$ we can use a Taylor approximation
%in \ref{infor1} with
%\begin{equation}
%\frac{M_{A}^2}{\mu_{0} \left(1-M_{A}^2\right)} \approx
%\frac{M_{A}^2}{\mu_{0}} \left(1+0.5 M_{A} \right) \approx
%\frac{M_{A}^2}{\mu_{0}}
%\end{equation}
the inertial force is a transformed part of the magnetic tension
force of the magnetohydrostatic field.
% (times the Alfv\'en Mach number squared).

We are now going to analyse the right hand side of the momentum or force equation
Eq.\,(\ref{euler1}), where we can identify the Lorentz force
by
%%%%%%%%%%%%%%%%%%%%%%%%%%%%%%%%%%%
\begin{eqnarray}
\vec j\times\vec B=\vec f_{L1} + \vec f_{L2} \, ,
\label{lorentz1}
\end{eqnarray}
%%%%%%%%%%%%%%%%%%%%%%%%%%%%%%%%%%%%%%%%%%%%%%%%%%%%
where 
\begin{eqnarray}
\vec f_{L1} &&= \frac{\vec j_{s}\times\vec B_{s}}{1-M_{A}^2}\label{lorentzf1}\\
\vec f_{L2} &&= -\frac{1}{\mu_{0}}\,\frac{M_{A} B_{s}^2}
{\left(1-M_{A}^2\right)^2}\,\vec\nabla M_{A} \label{lorentzf2}\, .
\end{eqnarray}
%%%%%%%%%%%%%%%%%%%%%%%%%%%%%%%%%%
Here we recognize that the initial Lorentz-force is enhanced by a factor
$1/(1-M_{A}^2)$ and changed by a term depending on the gradient of the
Alfv\'enic Mach number and the magnetic pressure.
But the the term with gradient of $M_A$ is also influenced strongly by the
strength of the Alfv\'en Mach number itself. If $M_A\stackrel{<}{\sim} 1$ the
term with the gradient of $M_A$ is enhanced by the $(1-M_{A}^2)^2$-term, having of course
even a stronger influence than the $(1-M_{A}^2)$-term in the nominator of the modified original
Lorentz-force $\vec f_{L1}$.

The second term on the right hand side of the momentum Eq.\,(\ref{euler1})
is given by the pressure force $\vec\nabla p_{s}$ transformed via
%({\bf hier noch ne Gleichungsreferenz von einem der alten Artikel!})
%%%%%%%%%%%%%%%%%%%%%%%%%%%%%%%%%%%%%%%%%%%%%%%%%%%%%%%%%%
\begin{eqnarray}
-\vec\nabla p   = && -\vec\nabla\left(p_{S}-\frac{1}{2\mu_{0}}\frac{M_{A}^2\vec B_{s}^2}{1-M_{A}^2}
\right)\\
= &&  - \vec\nabla p_{S} \label{statp1} \nonumber \\
&&  - \frac{M_{A}^2}{1-M_{A}^2}\,\vec j_{S}\times\vec B_{S}
\label{magp1} \nonumber \\
&& + \frac{1}{2\mu_{0}}\, B_{S}^2\,\vec\nabla\frac{M_{A}^2}{1-M_{A}^2}
\label{magp2} \nonumber \\
&&  + \frac{M_{A}^2}{\mu_{0}\left(1-M_{A}^2\right)}\left(\vec B_{s}
\cdot\vec\nabla\right)\vec B_{s} \, .
\end{eqnarray}
%%%%%%%%%%%%%%%%%%%%%%%%%%%%%%%%%%%%%%%%%%%%%%%%%%%%%%%%%%%%%%%%
Thus after final addition of the new pressure gradient to the new Lorentz-force
on the right hand side of the momentum equation Eq.\,(\ref{euler1}) the only
remaining force-term
is the net force, namely the modified magnetohydrostatic
tension force 
%%%%%%%%%%%%%%%%%%%%%%%%%%%%%%%%%%%%%%
\begin{equation}
\frac{1}{\mu_0}\,M_{A}^2/(1- M_{A}^2)
(\vec B_{S}\cdot\vec\nabla)\vec B_{S}\equiv\rho(\vec v\cdot\vec\nabla)\vec v \, .
\label{tension1}\end{equation}
%%%%%%%%%%%%%%%%%%%%%%%%%%%%%%%%%%%%%%%%%%%%%%%%%%%%%%%%
The remaining term
\begin{equation}
\vec\nabla p_{S}-\vec j_{S}\times\vec B_{S}
\end{equation}
%%%%%%%%%%%%%%%%%%%%%%%%%%%%%%%%%%%%%%%%%%%%%%%%%%%%%%%%%%
vanishes, because this describes the original static equilibrium.
As it can be recognized by the inertial force ($=$tension force)
in Eq.\,(\ref{tension1}), the expression can become extremely large in the
limit of $M_A\rightarrow 1$.

This implies that the modified pressure gradient consists of four terms:
the first is the static pressure, the second a term
completing the new Lorentz force to the old static one, the third term removes
the gradient forces concerning the Mach number gradient term in Eq.\,(\ref{lorentz1}),
and the fourth term displays the final inertia or net force term of the flow,
induced by the mapping method.

%Regarding this from the other point of view, namely to reduce on to the
%% ascribe to the mhs field
%static field, it implies that the new Lorentz-force
%has to consist of one term, compensating the negative part of the Bernoulli-pressure,
%depending on the gradient of the Alfv\'en Mach number. This term completes the
%new pressure gradient together with some part of the stagnation pressure that
%originates from the magnetic pressure gradient of the magnetohydrostatic field,
%Eq.\,(\ref{magp2}). 
%
%On the other hand, the remaining part of the Lorentz-force (second term of
%Eq.\,(\ref{lorentz1})),\lq modified\rq~by a term, depending on the Alfv\'en
%Mach number, can be summarized with the Lorentz-force term Eq.\,(\ref{magp1}),
%and therefore completes the static Lorentz-force.

\subsection{Comparison between quasi-static and stationary approach}

If the gradient of $M_A$ is very large and also $M_{A}$ has a value close to one, as explained in the
last section and following the last paragraph of the Lorentz-force Eq.\,(\ref{lorentz1}),
hence the complete Lorentz-force can become extremely large.
The situation in the quasi-static approach is different:
Although the current density grows extremely in the forming
thin current sheet, no strong Lorentz-forces occur,
due to the null sheet of the magnetic field, ignoring a small normal component to the sheet.
 
In the case of the stationary states even for flows with small or
moderate Alfv\'en Mach numbers, the magnetic flux density
and therefore the magnetic flux is not essentially enhanced, compared to the
initial MHS state, but the newly generated current density is able to increase in order
of magnitude. Additionally, filamentary or \lq fractal\rq~structures compared to the old
magnetohydrostatic current distribution are created, i.e. a double-structure evolves,
see \cite{2010AnGeo..28.1523N}, and also see Figs.\,(\ref{fig1}, \ref{fig2}).
The pressure gradient is enhanced
and has to be compensated by the Lorentz-force as the additional inertial force due
to the flow is of second order in $M_A$, see Eq.(\ref{weber2}) for small $M_{A}\ll 1$,
and therefore negligible, but not negligible for large $M_A$.
The only price one has to pay here
is the unchanged shape of the fieldlines, in contrast to the quasi-static approach,
and the formation of strong shear flows, i.e. small scale vortices, as we will
show in the next sections.

The additionally occuring
filamentary fine structures of the steady-state current sheets,
i.e. thin current sheets, are therefore similar to results found, e.g.,
in \cite{wiegelmann:etal95}.
These authors found that only for non-similarity solutions of the quasi-static equations
double-structure of the current density evolves. The evolution of a double-structure is necessary
for the occurence of thin current sheets.

However, the method of thin current sheet formation and structure, used in this paper,
is a different approach than
the quasi-static approach, e.g., see \cite{1982JGR....87.2263S} or \cite{wiegelmann:etal95}, where 
the pressure as boundary condition is applied and via the law of adiabatic change
is equivalent to magnetic flux transport and therefore {\it flux enhancement}. This is in contrast to
the approach of this paper, where at least for small Alfv\'en Mach numbers the flux is not essentially enhanced.

One interesting aspect of the velocity fields, being explained and calculated in
\cite{1988JGR....93.5922W} and \cite{1987JGR....92...95S},
is that the velocity component parallel to the magnetic field in the vicinity of separatrices is getting
very large. Hence also in the quasi-static regime, the flow parallel to the magnetic field
is of huge importance for the plasma dynamics.
%enhancement from the boundaries plays an important role
%in current sheet formation and structure.}

\section{Special cases}
\label{sec3}

To see how the transformation technique affects the initial MHS-equilibrium
we will now discuss and analyse some properties of specially chosen
MHS-equilibria and flows, i.e. Alfv\'en Mach numbers.

\subsection{Flows with global constant Alfv\'en Mach number}

\subsubsection{Potential field as initial static equilibrium}

The most simple cases of field-aligned incompressible MHD flows
seem to be represented by potential fields as initial
MHS-fields and the restriction to constant Alfv\'en Mach numbers. We now analyse some
combinations of these possibilities.
Discussing $j_{s}=0$ and $M_{A}=$const it can clearly be recognized in the modified
Lorentz force Eq.\,(\ref{lorentz1}) that
the magnetic field is similar to the old magnetohydrostatic field, i.e. the static
field does not change its relative field strength across the field lines.
The magnetic field stays potential, no currents form.
The gradient forces vanishes, i.e. $\vec\nabla p_{S}=\vec 0$,
there is no Lorentz-force, and only the constant
static pressure $p_{S}$ changes to an inverted shaped thermal pressure
$p$, reduced by a fraction of the magnetic pressure, depending  on $M_{A}$.
One needs large enough offsets in that case,
to avoid negative thermal pressures $p$.
The net force or inertial force from the flow can then be interpreted as
the gradient force of the plasma pressure in the magnetohydrostatic case
times $M_{A}^2/(1-M_{A}^2)$, which
for $\vec\nabla\times\vec B_{S}=0$ is identical to the magnetic tension force.
Without gradients of $M_{A}$ no additional forces can be generated.
\subsubsection{Initial force-free fields}
The Lorentz-force equation Eq.\,(\ref{lorentz1}) shows also clearly that
instead of the initially force-free state a force
is induced by the flow gradient term $\vec\nabla M_A$. All other aspects are similar to
the potential field case.

\subsubsection{Non force-free initial magnetohydrostatic equilibria}

With this trivial case that $j_{s}\neq 0$ and $M_{A}=$const, we get solutions
very similar to the initial MHS equilibria. The magnetic field and the
current are mapped by a constant factor $1/\sqrt{1-M_{A}^2}$, and therefore
the Lorentz force is enhanced by the constant factor $1/(1-M_{A}^2)$.
The current sheet structure is conserved, and therefore basically these solutions
represent similarity solutions, as all terms, including gradients of the Mach number,
vanish.
%\\
%$\rightarrow$  Discuss potential field with shear component! How is $B_{zS}$ mapped?
%Effects? Conclusions!
%
\subsection{Flows with non-constant Alfv\'en Mach number profile}
The generic case of plasma flows is that the Mach number changes
from one field line to the other and the transformation equations
are non-linear. This leads to additional nonlinear effects as discussed in the
following.
\subsubsection{Initial potential field}
The special case of $j_{s}=0$ and $M_{A}\neq$const is simple, but very interesting
for easy modelling: one can choose a potential, e.g.
asymptotical homogeneous field and map it with
multiple current sheets, as done in \cite{2006A&A...454..797N}.
The effect on the forces is that the Bernoulli pressure force within
the new pressure force, i.e. $-\vec\nabla p=-\vec\nabla(p_{S}-\rho\vec v^2/2)$
must be compensated by an additional Lorentz-force. As the $\vec f_{L1}$-term in
Eq.\,(\ref{lorentz1}) is zero, as $j_{S}=0$, only the force-term  $\vec f_{L1}$
in Eq.\,(\ref{lorentz1}) can do that job.
The reason is that only the gradient of the Alfv\'en Mach number can
generate currents and then induce additional Lorentz-forces.
The net force or inertial force originating from the flow can then be interpreted as
the gradient force of the magnetohydrostatic pressure times $M_{A}^2/(1-M_{A}^2)$.
The fact that $M_{A}$ is non-constant produces currents perpendicular to the
magnetic field, repeating and keeping in mind that,
%%%%%%%%%%%%%%%%%%%%%%%%%%%%%%%%%%%%%%%
\begin{eqnarray}
\mu_{0}\vec j= && M_{A}\,\frac{\vec\nabla M_{A}\times\vec B_{S}}
{\left(1-M_{A}^2\right)^{3/2}}+
\frac{\vec\nabla\times\vec B_{S}}{\left(1-M_{A}^2\right)^{1/2}}\nonumber\\
\nonumber\\
\stackrel{\textrm{\tiny here}}{=} &&
M_{A}\,\frac{\vec\nabla M_{A}\times\vec B_{S}}{\left(1-M_{A}^2\right)^{3/2}
} \, . \nonumber\\
\label{current1}
\end{eqnarray}
%%%%%%%%%%%%%%%%%%%%%%%%%%%%%%%%%%%%%%
The equation reflects that in the this case the generated currents
are perpendicular to the magnetic field but also to the direction in which
the flow gradient is oriented.

\subsubsection{Initial force-free fields}

While the first term in the first line of Eq.\,(\ref{current1}) provides only
perpendicular currents with respect to the magnetic field, if the Alfv\'en Mach number has
strong gradients, the effect of the second term is only to enhance the existing
magnetohydrostatic force-free current, being parallel to the magnetic field, writing
%%%%%%%%%%%%%%%%%%%%%%%%%%%%%%%%%%%%%%%%%%%%%%%%%%%%%%%
\begin{eqnarray}
\mu_{0}\vec j= && M_{A}\,\frac{\vec\nabla M_{A}\times\vec B_{S}}
{\left(1-M_{A}^2\right)^{3/2}}+
\frac{\vec\nabla\times\vec B_{S}}{\left(1-M_{A}^2\right)^{1/2}}\nonumber\\
= && M_{A}\,\frac{\vec\nabla M_{A}\times\vec B_{S}}
{\left(1-M_{A}^2\right)^{3/2}}+
\frac{\alpha\vec B_{S}}{\left(1-M_{A}^2\right)^{1/2}}\, ,
\label{current1b}
\end{eqnarray}
%%%%%%%%%%%%%%%%%%%%%%%%%%%%%%5
where $\vec\nabla\times\vec B_{S}=\alpha\vec B_{S}$. This implies that a former force-free field without
flows, now develops components of the current, producing forces, pointing into the direction of
$\vec\nabla M_{A}$, see modified Lorentz-force in Eq.\,(\ref{lorentz1}).

\section{Connection between vortices and currents in field-aligned, subalfv\'enic,
stationary ideal MHD}
\label{sec4}
Even if the notion of \lq turbulent\rq~and \lq ideal\rq~reconnection, see,
e.g. \cite{2006PhyD..223...82E} or other references, are based on discontinuous solutions,
also including a specific angle between vortex sheet and current sheet, the solutions
presented here, represents a powerful method to find the limit from continuous,
namely well known MHS-solutions with smooth current sheets, to almost
discontinuous solutions, i.e. sufficiently small subscales of concentrated
 and strong current sheets.
It is obvious that for $\vec v\parallel \vec B$ the \lq curls\rq~of both vector fields,
i.e. their vortices, are directly connected, of course not necessarily parallel
\footnote{For global constant $M_{A}/\sqrt{\rho}$ they are strictly parallel.}
%%%%%%%%%%%%%%%%%%%%%%%%%%%%%
\begin{eqnarray}
\vec v = \frac{M_{A}\,\vec B}{\sqrt{\mu_{0}\rho}}:=\lambda\,\vec B
\,\,\,\,\,\,\,\Rightarrow\,\,\,\,\,\,\, \vec\nabla\times\vec v=  \vec\nabla\lambda\times\vec B
+ \lambda \vec\nabla\times\vec B \,\, .
\label{vortex1}
\end{eqnarray}
%%%%%%%%%%%%%%%%%%%%%%%%%%%%%%%%%%%%%%%%%%%
As $\vec\nabla\times\vec B$ can be split into one perpendicular component
$\vec j_{\perp}$ with respect to the magnetic field and one parallel component,
$\vec j_{\parallel}$, we can rewrite the curl of the velocity
%%%%%%%%%%%%%%%%%%%%%%%%%%%%%%%%%%%%%%%%%%%%%%%%%%%%%%%%%%%%%%%
\begin{eqnarray}
\vec\nabla\times\vec v= \vec\nabla\lambda\times\vec B +\frac{\lambda}{\mu_{0}}
\left(\vec j_{\parallel}+\vec j_{\perp}\right)\, \,\, .
\end{eqnarray}
%%%%%%%%%%%%%%%%%%%%%%%%%%%%%%%%%%%%%%%%%%%%%%%%%%%%%%%%%%%%%%%%%
This implies that %only in the case that
\begin{eqnarray}
\vec B\cdot(\vec\nabla\times\vec B)=0 \quad\Rightarrow\quad \vec j_{\parallel}=\vec 0
\label{vanpacurrent}
\end{eqnarray}
%%%%%%%%%%%%%%%%%%%%%%%%%%%%%%%%%%%%%%%%
is a necessary condition for a MHD configuration with a non-vanishing current but
without vortices in 3D (or 2D).
This is necessary as the term with the gradient of $\lambda$, being perpendicular to
the magnetic field, cannot compensate the term with $\vec j_{\parallel}$.
For a 2.5D configuration the parallel part of the current can only
vanish identically if everywhere $(\vec\nabla A)^2$ can be expressed as
a function of $A$. Thus, a physical situation with currents, but without
vortices it is necessary that the parallel current vanishes. We will give an
example, to be precise the sufficient criterion in 2D (Eq.\,(\ref{maro})).
%Taking the curl on both sides of Eq.\,(\ref{vortex1})
%%%%%%%%%%%%%%%%%%%%%%%%%%%%%%%%%%%%%%%%%%%%%%%%%%%%%
%\begin{eqnarray}
%\vec\omega=\vec\nabla\times\vec v = \vec\nabla\lambda\times\vec B
%+ \lambda\,\vec\nabla\times\vec B
%\end{eqnarray}
%%%%%%%%%%%%%%%%%%%%%%%%%%%%%%%%%%%%%%%%%%%%%%%%%%%%%
%and scalar multiplication with $\vec\nabla\lambda$ and $\vec B$
%
%\begin{eqnarray}
%&&\left(\omega-\frac{\lambda}{\mu_{0}}\vec j\right)\cdot\vec\nabla\lambda=0\\
%&&\left(\omega-\frac{\lambda}{\mu_{0}}\vec j\right)\cdot\vec B=0
%\end{eqnarray}
%%%%%%%%%%%%%%%%%%%%%%%%%%%%%%%%%%%%%%%%%%%%%%%%%%%%%
%\begin{eqnarray}
%&& \left(\vec\nabla\times\vec\omega-\frac{\lambda}{\mu_{0}}\frac{M_{A}\,\vec B}{\sqrt{\mu_{0}\rho}}
%\right)\cdot\vec\nabla\frac{M_{A}\,\vec B}{\sqrt{\mu_{0}\rho}} =
%0\\
%&& \left(\vec\nabla\times\vec v\right)\cdot\vec B
%\end{eqnarray}
%%%%%%%%%%%%%%%%%%%%%%%%%%%%%%%%%%%%%%%%%%%%%%%%%%%%

%
With the additional assumptions that the flow is incompressible we can simplify
the relations (\ref{vortex1}), getting the vorticity or vortex (strength) $\vec\Omega$,
%%%%%%%%%%%%%%%%%%%%%%%%%%%%%%%%%%%%%%%%%%%
\begin{eqnarray}
 \vec\Omega &&=\vec\nabla\times
\left(
\frac{M_{A}\vec B}{\sqrt{\mu_{0}\rho}}
\right)
%\nonumber\\
= -\frac{M_{A}\vec\nabla\rho\times\vec B}{2\left( \mu_{0}\rho^{3}\right)^{1/2}} + \frac{
\vec\nabla\times\left(
M_{A}\vec B\right)}{\sqrt{\rho}}
\nonumber\\
&&= \frac{M_{A}\vec B_{S}\times\vec\nabla\rho}{2\sqrt{\mu_{0}}\,\left(1-M_{A}^2\right)^{1/2} \rho^{3/2}}
+\frac{\vec\nabla M_{A}\times\vec B_{S}}
{\sqrt{\mu_{0}\rho}\left(1-M_{A}^2\right)^{3/2}} \nonumber\\
&&\quad + \frac{M_{A} \vec\nabla\times\vec B_{S}}
{\sqrt{\mu_{0}\rho}\,\left(1-M_{A}^2\right)^{1/2}}\, .\nonumber\\
\label{vortex2}
\end{eqnarray}
%%%%%%%%%%%%%%%%%%%%%%%%%%%%%%%%%%%%%%%%%%%%%%%%%%%%%%%%%%%%%%%%%%%%%%%
For potential fields ($\vec\nabla\times\vec B_S=\vec 0$) the vorticity
only has components perpendicular to the magnetic field, if $M_{A}$ is constant,
then the vorticity is only non-zero, if the density is not completely constant.
Strong vortex sheets only occur if the density gradient is strong enough.

In the case of force-free fields there are nonvanishing components of the vorticity, even
if $M_{A}$ is constant, while the current remains field-aligned.
The second term in the last equation also enables to generate a component
of the vorticity perpendicular to the magnetic field in the case
$\vec\nabla M_{A}\neq\vec 0$, but it is, of course, also perpendicular to the current. 

In the nonlinear case with small $M_A=$const, the vortex can be large, while the current is not
extremely enhanced by the flow.
Necessary for $\vec\Omega=\vec 0$, e.g., in  2D with
$\vec\nabla\times\vec B_{S}=\vec 0$ is, with apostrophes denoting
derivatives with respect to the flux function $A$,
%%%%%%%%%%%%%%%%%%%%%%%%%%%%%%%%%%%%%%%%%%55
\begin{eqnarray}
&& \frac{\vec\nabla M_{A}\times\vec B_{S}}{\sqrt{\left(\mu_{0}\rho\right)
\left(1-M_{A}^2\right)}} =\frac{M_{A}}{\mu_{0}}\, \frac{\vec\nabla\rho\times\vec B}{2\rho^{3/2}}\nonumber\\
&& \frac{M_{A}'(A)\left(\vec\nabla A\times\vec B_{S}\right)}{
\sqrt{\left(\mu_{0}\rho\right)}\left(1-M_{A}^2\right)^{3/2}}
= \frac{\rho'\left(A\right) M_{A}\left(A\right)
\left(\vec\nabla A\times\vec B_{S}\right)}{2\rho^{3/2}\sqrt{\mu_{0}}
\left(1-M_{A}^2\right)^{1/2}}\nonumber\\
\Rightarrow && \frac{M_{A}'\left(A\right)}{M_{A}\left(A\right)
\left(1-M_{A}^2\left(A\right)\right)}=\frac{\rho'\left(A\right)}{\rho\left(A\right)}\nonumber\\
\Rightarrow && \rho = \rho_{0} \left(\frac{M_{A}^2}{ 1-M_{A}^2}\right)^{1/2}\, .\nonumber\\
\label{maro}\,\,\, 
\end{eqnarray}
%%%%%%%%%%%%%%%%%%%%%%%%%%%%%%%%
The density $\rho$ and the Alfv\'en Mach number are {\it usually} independently conserved
quantities along the field lines, i.e. both $\rho=\rho(A)$ and $M_{A}=M_{A}(A)$ can be
freely choosen.
One recognizes that this cannot be fulfilled, if the vorticity should vanish.
Here, either the density is a specified function of
the Alfv\'en Mach number, i.e. $\rho=\rho(M_{A}(A))$,
or vice versa $M_{A}=M_{A}(\rho(A))$.
% free integrals
%of the magnetic field
If $\vec\nabla\times\vec B_{S}\neq\vec 0$ than a solution is only possible if again
relation Eq.\,(\ref{vanpacurrent}) holds, i.e. the current must be perpendicular
to the magnetic field.
Concerning the current equation, Eq.\,(\ref{current1}) it can be recognized that
this necessary condition Eq.\,(\ref{vanpacurrent}) must also be fulfilled, if one wants the
steady-state-current to vanish.
In the 2D case the current distribution and the magnetic field
are then basically 1D. This could, e.g., lead either to field lines being circles or straight
field lines in 2D.

In general (3D), if  $M_{A}$ is non-constant and $\vec B_{S}$ is a non-potential field, then
the current can only vanish if $\vec j_{S\parallel}=\vec 0$.

If $M_{A}$ is constant, or its gradients can be neglected, and the corresponding
magnetohydrostatic field is potential it is possible to have no currents, or current
sheet, but with the gradient term of the density it is possible to generate a vortex
sheet in Eq.(\ref{vortex2}) at the same time.

\subsection{Sequence of equilibria towards almost
magnetohydrostatic configurations}

To analyse the connection between current and vortex sheets
without loss of generality in 2D, we search for
an ansatz for the Alfv\'enic Mach number sheet.
We use an ansatz where the Alfv\'en Mach number $M_{A}$ is depending
on one flux function, but basically this view can be transferred
to the problem in 3D with two flux functions, i.e. Euler potentials.
We present the principle how to construct a
series of peaked function for the Alfv\'en Mach number, sharply
separating regions with flow from regions with almost no flow ($M_{A}\ll 1$).
The Alfv\'en Mach number should therefore be represented
by a function series depending on a smallness parameter $\varepsilon\neq 0$, which
tends to zero, to mimic an
approximately {\it static} equilibrium. This has relevance for a lot of situations in
which the complete or large part of the domain have a basically quasi-magnetohydrostatic
character, with some regions showing almost no flows along the field line, e.g., \cite{2005ESASP.596E..12N}
%Neukirch 2005ESASP.600E.31N, Forbes and Priest 1995,
and \cite{1999ESASP.448..871R}. 
The relevance for the case $M_{A}\ll 1$ is valid especially in the case of a low $\beta$--plasma, as
%%%%%%%%%%%%%%%%%%%%%%%%%%%%%%%%%%%%%%%%%%%%%%%%%%%%%%%%%%%%%%%%%%%
\begin{equation}
1\gg \beta=\frac{p}{\vec B^2/(2\mu_{0})}=\frac{p/\rho}{\vec B^2/(2\mu_{0} \rho)}
=\frac{2}{\gamma}\frac{M_{A}^2}{M_{S}^2}
\end{equation}
%%%%%%%%%%%%%%%%%%%%%%%%%%%%%%%%%%%%%%%%%%%%%%%%%%%%%%%%%%%%%%%%%%%
$\gamma$ being the polytropic exponent and $M_{S}$ the usual Mach number, i.e.
%%%%%%%%%%%%%%%%%%%%%%%%%%%%%%%%%%%%%%%%%%%%%%%%%%%%%%%%%%%%%%%%%%%%%%%%%
\begin{equation}
\frac{p}{\rho^{\gamma}}={\rm const} \, ,
\quad v_{S}^2=\frac{dp}{d\rho}\, , \quad M_{S}^2=\frac{v^2}{v_{S}^2}
\end{equation}
%%%%%%%%%%%%%%%%%%%%%%%%%%%%%%%%%%%%%%%%%%%%%%%%%%%%%%%%%%%%%%%%%%%%%%%%%%%%
Approximate incompressibility is plausible for Mach numbers $M_{S}\ll 1$, and hence
the condition $M_A\ll M_{S}\ll 1$, to guarantee a low $\beta$--plasma, is even more plausible.
%%Romeou \& Neukirch. % 2002Jastp64, etc.).
%{\bf The procedure of a small Alfv\'en Mach number should 
%not be misunderstood as a strict 
%mathematical
%limes,} because we need both, flow {\it and} field but it shows the tendencies, namely how for
%different partitions of kinetic and magnetic pressure vortices and currents behave.
In fact, $M_A\stackrel{<}{\sim} 1$ is possible for regions in which the plasma $\beta$ is
much larger, e.g., in the vicinity of (degenerated) separatrices, being null sheets of the magnetic field
, or regions where the magnetic field in general is weaker.

Taking a normalized flux function $A$ and a scaling factor
$\varepsilon$ into account, a very general ansatz for $M_{A}$ can be written as
the following sequence of functions
%%%%%%%%%%%%%%%%%%%%%%%%%%%%%%%%%%%%%%
\begin{eqnarray}
M_{A} &=& \pm |M_{A, max}|\,\varepsilon^{n}\left[\frac{\pm 1 +
\tanh\left(\frac{A}{\varepsilon}\right)
}{2}\right]
\label{mach1}
\\
\lim_{\varepsilon\rightarrow 0} M_{A} &=& 0\label{mach2}\\
\frac{d M_{A}(A)}{dA} &\equiv & M_{A}'(A) < \infty\label{mach3}\\
\Rightarrow\quad M_{A}' &=& \varepsilon^{n-1}\,\cosh^{-2}{\left(\frac{A}{\varepsilon}\right)}
\label{mach4}\\
M_{A} M_{A}' &=& \varepsilon^{2n-1}\left[\cosh^{-2}{\left(\frac{A}{\varepsilon}\right)}
+\frac{\sinh\left(\frac{A}{\varepsilon}\right)}{\cosh^{3}{\left(\frac{A}{\varepsilon}
\right)}}\right]\label{mach5}
\end{eqnarray}
%%%%%%%%%%%%%%%%%%%%%%%%%%%%%%%%%%%%%%%%
i.e., the terms with $M_{A}$ and $M_{A}'$
are bounded for $\varepsilon\neq 0$. A divergence of these terms and therefore the
parametric formation of a vortex sheet can be recognized in the second row of
the vortex equation, Eq.\,(\ref{vortex2}) if we consider $\lim\varepsilon\rightarrow 0$,
and, of course, taking into account that $\vec\nabla M_{A}(A)=M_{A}'(A)\vec\nabla A$.
For small Alfv\'en Mach numbers, even if we do not only take potential fields, i.e.
$\vec\nabla\times\vec B_{S}=\vec 0$, into account, the dominant and important terms
are that term in the vortex and current equations , that include at least the
$\vec\nabla M_{A}$-term.
We can distinguish different regimes for the exponent $n$, which are listed
in the following:

\begin{itemize}

\item (i) For  $n > 1$, the current sheet and the vortex
sheet contract with $1/\varepsilon$ in Eq.\,(\ref{mach5}),
but their amplitudes converge to zero, as can be seen by
taking the limit for Eq.(\ref{mach5}) for $\varepsilon\rightarrow 0$.
The reason is that the $M_{A}\, M_{A}'$ term converges with
$\varepsilon^{2n-1}=\varepsilon^{2(1+|\delta|)-1}=\varepsilon^{1+|\delta|}$ to zero.

\item (ii) For $n=1$ the vortex sheet and the current contract, the amplitude of
the vortex sheet stays finite, the current sheet amplitude converges to zero,
also due to the $M_{A}\, M_{A}'$-convergence, reading $\propto \varepsilon$.

\item (iii) In the range $1>n>\frac{1}{2}$ the vortex sheet and the current contract, the
vortex sheet amplitude diverges with $\varepsilon^{-(1-n)}$ for
$\varepsilon\rightarrow 0$, the amplitude of the current sheet converges to zero
with $M_{A}\, M_{A}' \propto \varepsilon^{|\delta|}$.

\item (iv) For $n=\frac{1}{2}$ the vortex sheet and the current contract, the vortex
sheet diverges with $\varepsilon^{-\frac{1}{2}}$, the amplitude of the current
sheet stays finite, converging with $M_{A}\, M_{A}' \propto\varepsilon^{0}$.

\item (v) For $0\leq n<\frac{1}{2}$ the vortex sheet and the current sheet contract, the vortex
sheet (with $\propto \varepsilon^{-2|\delta|}\,\, , \frac{1}{2} \geq |\delta|> 0$ )
and the current sheet amplitude diverge
$M_{A}\, M_{A}' \propto\varepsilon^{-\frac{1}{2}|\delta|}$;
for $n=0$ the Alfv\'en Mach number
stays finite and therefore cannot converge to a quasi static case. We therefore do not take
values of $n<0$ into consideration.

\end{itemize}

%\subsubsection{Preliminary discussion}

These different cases imply that only for a special chosen funcion of the
Alfv\'en Mach number current and vortex sheets simultaneously appear, namely for
the case of the exponent $n<1/2$. For the other cases it is at least necessary that
$n<1$ to generate a singular vortex sheet in the limit $\varepsilon\rightarrow 0$, but
without getting a current singularity. Case (v) guarantees the existence of
both, a current sheet and a vortex sheet.

We did not discuss the influence of the $\vec\nabla\rho$-term in the vortex-equation,
Eq.\,(\ref{vortex2}), very intensively. The influence can also be very important,
although it is with a $\propto M_{A} M_{A}'$-dependence of the same strength and
quality as the dependence of the current density.

\subsection{General view on the connection between currents and vortices}

%The current sheet contracts/is getting thinner for $n=\frac{1}{2}$ but stays finite
%and constant(?), but the vortex/vortices diverge(s)
%guarantees that the functions
%\\
%
%\vspace{0.5cm}
%
To clarify the structure of this connection
between vortex sheets and current sheets from a different view,
we introduce the abbreviations
$\vec G_{A}=(\vec\nabla M_{A}\times\vec B_{S})/
(1-M_{A}^2)^{3/2}$ and $\vec J_{S}=\vec j_{S}/(1-M_{A}^2)^{1/2}$
and can rewrite the vortex-equation Eq.\,(\ref{vortex2}) and the current-equation
Eq.\,(\ref{current1})
%%%%%%%%%%%%%%%%%%%%%%%%%%%%%%%%%%%%%%%%%
\begin{eqnarray}
&& \vec j=\frac{M_{A}}{\mu_{0}} \vec G_{A} + \vec J_{S}\label{connex1a}\\
&& \vec\Omega= \frac{1}{\sqrt{\rho}}\,\vec G_{A} +
\frac{M_{A}}{\mu_{0}\sqrt{\rho}}\,\vec J_{S}-\frac{M_{A}\,\rho^{-\frac{3}{2}}}{2\sqrt{\mu_{0}}}
\vec\nabla\rho\times\vec B \label{connex1b}\\
\Rightarrow &&
\vec\Omega=\frac{\rho^{-\frac{3}{2}}}{2}
\frac{M_{A}}{\mu_{0}}\vec\nabla\rho\times\vec B  + \frac{1}{\sqrt{\rho}}\vec G_{A} +
\frac{M_{A}}{\mu_{0}\rho}\left(\vec j-\frac{M_{A}}{\mu_{0}}\vec G_{A}\right)\nonumber\\
\label{connex2}
\end{eqnarray}
%%%%%%%%%%%%%%%%%%%%%%%%%%%%%%%%%%%%%%%%%%%%%

The terms of order $O(M_A^2)$ can be neglected for small values of $M_A$. Then
one can recognize that the most important term, the zeroth order in $M_A$, is the $\vec G_A$-term,
followed by the gradient term of the density and the term with the current density
as terms of first order in $M_A$. 

But even for the general case $M_A\stackrel{<}{\sim} 1$ the $\vec G_A$-term has the strongest
influence on the generation of vorticity, as it scales only with $\rho^{-1/2}$ in contrast to the
$\vec j$-term with a $1/\rho$-dependence or the $\rho^{-3/2}$-dependence of the $\vec\nabla\rho$-term.
Only if the scale of the density gradient
is much smaller than that of the Alfv\'en Mach number the gradient term of the density can dominate.
The $\vec G_A$-term is also favoured, as for values of $M_A$ close to one the $(1-M_{A}^{2})^{3/2}$-dependence
enhances the influence of the $\vec\nabla M_{A}$-term on the vorticity stronger,
than the $(1-M_{A}^{2})^{1/2}$-dependence of all other terms in the vorticity Eq.\,(\ref{connex2}).
The nature of this transformation technique shows that these steady-state current-vortex
structures are related to static equilibria, and therefore the vortex structure
{\it and its scales} governs the structure of the currents and current sheets.

%\section{HEADING}
%TEXT
%
%\subsection{HEADING}
%TEXT
%
%\subsubsection{HEADING}
%TEXT

%%%%%%%%%%%%%%%%%%%%%%%%%%%%%%%%%%%%%%%
%Bilder                               %
%%%%%%%%%%%%%%%%%%%%%%%%%%%%%%%%%%%%%%%
\section{Examples}
\label{sec5}

\begin{figure*}
\vspace*{2mm}
\begin{center}
\mbox{\includegraphics[width=6cm]{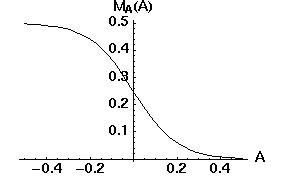}}
\includegraphics[width=6cm]{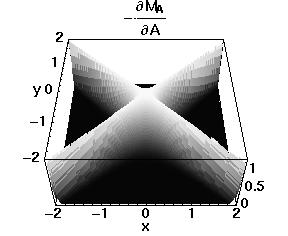}
\mbox{\includegraphics[width=6cm]{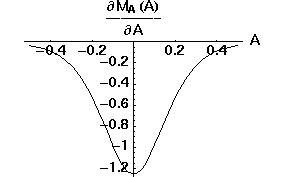}
\includegraphics[width=6cm]{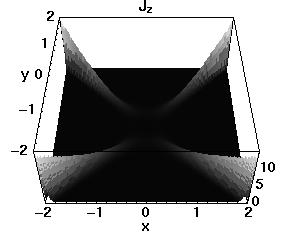}}
\mbox{\includegraphics[width=6cm]{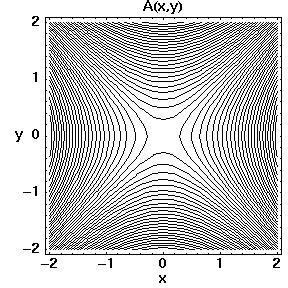}
\includegraphics[width=6cm]{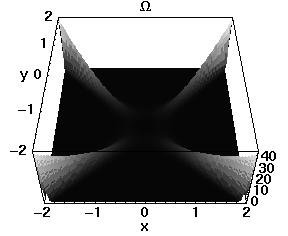}}
\mbox{\includegraphics[width=6cm]{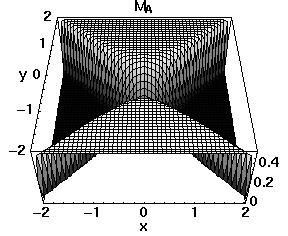}
\includegraphics[width=6cm]{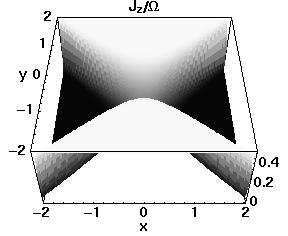}}
\end{center}
\caption{Transformation from an initial potential field
with a smooth Mach-number profile (see equation \ref{eq_ma}
with $M_{\rm A max}=0.5$ and $d=0.2$.)}
\label{fig1}
\end{figure*}

\begin{figure*}
\vspace*{2mm}
\begin{center}
\mbox{\includegraphics[width=6cm]{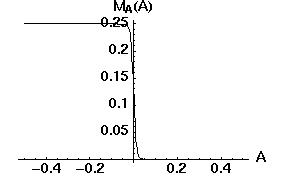}
\includegraphics[width=6cm]{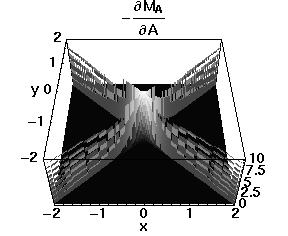}}
\mbox{\includegraphics[width=6cm]{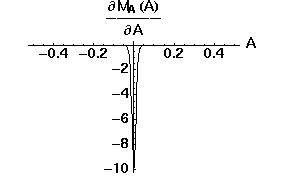}
\includegraphics[width=6cm]{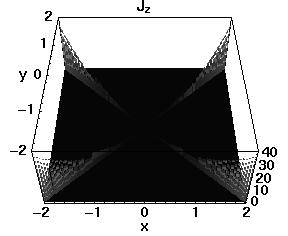}}
\mbox{\includegraphics[width=6cm]{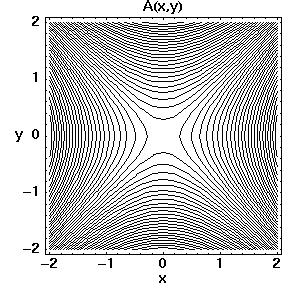}
\includegraphics[width=6cm]{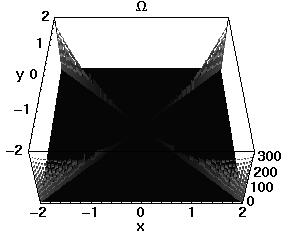}}
\mbox{\includegraphics[width=6cm]{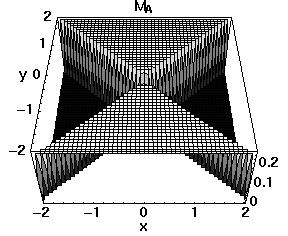}
\includegraphics[width=6cm]{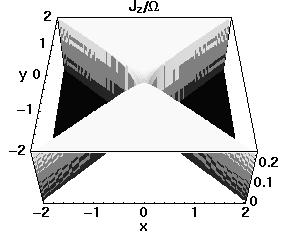}}
\end{center}
\caption{Transformation from an initial potential field,
similar as in figure \ref{fig1}, but with
a much steeper Mach-number profile (see equation \ref{eq_ma}
with $M_{\rm A max}=0.25$ and $d=0.2/16$.)}
\label{fig2}
\end{figure*}
As an example we investigate an initial 2D-equilibrium
in the form
\begin{equation}
A(x,y)=x^2+ (j_z/2-1) y^2,
\end{equation}
which reduces to a potential field solution for a vanishing
current $j_z$ as shown in figure \ref{fig1} panel $A(x,y)$.
We describe a Mach-number-profile in the form
\begin{equation}
M_A(A)=\frac{M_{\rm A \, max}}{2} \left(1-\tanh \frac{A}{d} \right),
\label{eq_ma}
\end{equation}
where $M_{\rm A \, max}=0.5$ and $d=0.2$ are free parameters,
see top-left panels \ref{fig1} showing $M_A(A)$ and $\frac{\partial M_A(A)}{\partial A}$
for the profile and its derivative, respectively. This introduces
a plasma flow on some field lines and a moderate flow gradient
perpendicular to the x-point separatrices as shown in the top-right
panels $M_A$ and $\frac{\partial M_A}{\partial A}$. As a consequence
the initial current free equilibrium develops a current and
vortex-sheet aligned with the separatrices (middle-right panels $J_z$ and
$\Omega$, respectively). The bottom-right panel $\frac{J_z}{\Omega}$ shows the ratio
of these quantities.

In figure \ref{fig2} we use the same mathematical form, but provide
a significantly steeper flow gradient $d=0.2/16$ along the separatrices , but
a lower maximum flow velocity $M_{\rm A \, max}=0.25$. The panels in figure
\ref{fig2} have the same meaning as in figure \ref{fig1}. We observe,
that both the current sheet and vortex sheet increase in magnitude,
but the vortex-sheet is stronger affected, as the ratio $\frac{J_z}{\Omega}$
shows.

\conclusions
\label{sec6}

Numerical incompressible time dependent
adaptive MHD-simulations by \cite{1998PhPl....5.2544G,2000PhRvL..84.4850G},
show that vortex-sheets align with current sheets. Although it is not clear why
this alignment occurs within the approach of these simulations,
we could show that for steady-state, incompressible field-aligned
flows this alignment appears naturally in a non-trivial way: even if the Alfv\'en Mach number
is non-constant, and non-trivial terms appear, resulting from the used non-canonical transformation
mechanism, the resulting microscopic or thin structure and decrease of the amplitude of current and vortex
strength show a clear alignment. The reason is that in both representations of current and vorticity
the gradient of the the Alfv\'en Mach number $M_A$ generates the thin structure of vortex and current sheet.

The larger the gradient, i.e. the smaller the lengthscale, the stronger the vorticity and the current.
The difference is only that the current generating term is multiplied by $M_A$, i.e. in the case
of approximative magnetohydrostatics $M_A\ll 1$ the vortex is stronger enhanced than the current.
As the scale for the flow, inducing the current sheet, must be very small for small
Alfv\'en Mach numbers, the solution can show strong vortex sheets but not so strong current sheets.

It is also noteworthy that the function $\rho$ can be chosen freely as a function 
of the Euler-potentials in the frame of this transformation
technique.
The free choice leads to strong influence of the vorticity on the density gradient. This may be
important for regions across magnetic boundaries, e.g. helmet streamer regions, the heliopause or
other astropauses, the magnetopause etc. where inner and
outer field line and plasma regions are separated by separatrices or magnetopauses, separating
regions of higher density from regions of lower density. Thus it is very difficult to get current
sheets without vortex sheets, as the used non-canonical transformation method automatically has
to produce shear flows, to create filamentary strucured current sheets. These shear flows in fact imply
vortex-sheets, hence generating current sheets.

On the other hand for special configurations it is also possible to create a current sheet with a
vanishing vortex strength, in 2D, where one {\it necessary} condition is automatically fulfilled, namely
that the parallel current with respect to the magnetic field vanishes, i.e. $\vec j_{\parallel}=\vec 0$.
But a special density distribution must also be included to be {\it sufficient}: normally the both free and
independent integrals $M_{A}$ and $\rho$ are then necessarily dependent functions of each other in
2D, i.e. $\rho=\rho(M_{A}(A))$.

The velocity fields calculated from the quasi-static approach, see \cite{1987JGR....92...95S}
and \cite{1988JGR....93.5922W} show high bulk velocities parallel to the magnetic field, similar to
the pure field-aligned flow of our presented calculations. There is also an analogy
to the fine structures of current sheets in both approaches, namely the filamentary structure formation,
i.e., the formation of a peaking, almost singular current, in time-dependent, quasi-static theory and
the filamentary structure formation of the current sheet due to small scale gradients
of Alfv\'en Mach number of our stationary approach.
%The enhanced vorticity and current are important for anomalous dissipation, e.g. therefore for
%coronal heating \cite{2003Natur.425..692S}}. {\bf (REFERENCES!!??), Craig paper
%mit viscosity, photospheric vortex motions!}, see % \cite{1981ITPS....9...21A}
The used technique to calculate exact solutions of ideal MHD from known magnetohydrostatic solutions
is very important to understand the structure of current-vortex sheets also regarding
eruptive space plasma processes like magnetic reconnection and following
magnetospheric substorms, eruptive flares or coronal mass ejections.
Even if anomalous viscosity is damped or minimized \cite{1981ITPS....9...21A}
and shear flows in specific situations lower the reconnection
rate \cite{1989BAICz..40...23K} it is important to generate smaller scale dissipation, see e.g.,
\cite{2003PhPl...10.4661B}.

\lq Flare observations should be used to
investigate whether the large-scale vortical flows, required to
sustain the maximum viscous dissipation rate, are indeed present
in the flaring solar corona.\rq, as theoretical investigations on steady reconnection and dissipation
show, see, \cite{2009A&A...501..755C}.

Hence we expect that in the vicinity of current sheets close to separatrices,
it is possible to observe such strongly sheared flows. From that point of view it would be interesting
to search for small scale shear flows in the vicinity of thin current sheets to compare this with
initial configurations in active or eruptive regions in the solar atmosphere or in magnetospheres
before magnetic (sub)storms. The involved scales may admittedly be extremely small.

It is also noteworthy that in adaptive mesh refinement analyses of magnetic reconnection, the
\lq fractal\rq~or better filamentary structure
of current sheets can also be found and plays a very important role in fragmented reconnection, i.e. current-layer
fragmentation, see e.g. \cite{2011ApJ...737...24B} and \cite{2010AdSpR..45...10B}.

It would be of great interest to investigate steady-state current-vortex sheets for general
field-aligned flows.

\begin{acknowledgements}
D.H.N. acknowledges support from a visitor grant of the Max-Planck-Institute in Katlenburg-Lindau.
%and from the Astronomical Institute AV \v{C}R in Ond\v{r}ejov.
\end{acknowledgements}

\bibliography{dieter}
\bibliographystyle{copernicus}

\end{document}